\documentstyle[aps,prl,preprint]{revtex}
\begin{document}
\title
{Phase Transition, Longitudinal Spin Fluctuations
 and Scaling in a Two-Layer Antiferromagnet}
\author{Andrey V. Chubukov$^*$ and Dirk K. Morr}
\address
{Department of Physics, University of Wisconsin-Madison,
1150 University ave., Madison, WI 53706\\
$^*$ P.L. Kapitza Institute for Physical Problems, Moscow, Russia}
\date{today}
\maketitle
\begin{abstract}
We consider a two-layer Heisenberg
antiferromagnet which can be either in the N\'{e}el-ordered or in the
disordered
phase at $T=0$, depending on the ratio of the intra- and interlayer
exchange constants.
We reduce the problem to an interacting Bose-gas
and  study the sublattice magnetization and
the transverse susceptibility in the
ordered phase, and the spectrum of quasiparticle excitations in both phases.
 We compare the results with the spin-wave theory and
argue that the longitudinal spin
fluctuations, which are not included in the spin-wave description,
 are small at vanishing coupling between the layers,
but increase as the system approaches the
transition point.
 We also compute the uniform
susceptibility at the critical point to order $O(T^2)$, and
show that the corrections to scaling
are numerically small, and
 the linear behavior of $\chi_u$ extends to high
temperatures. This is consistent with the results of the
recent Monte-Carlo simulations by Sandvik and Scalapino.
\end{abstract}
\narrowtext

\section{Introduction}

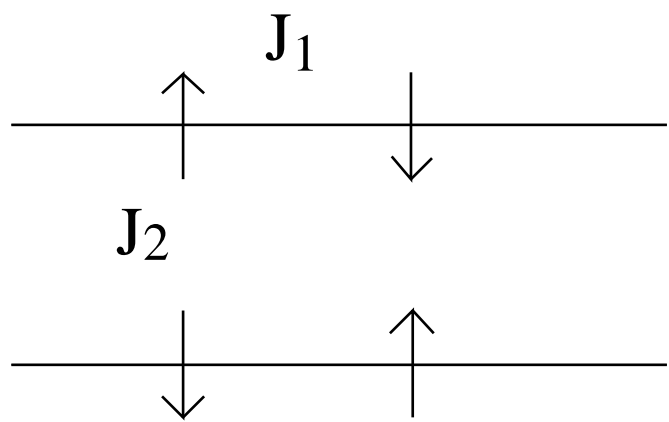
\begin{figure}
\caption{The system under consideration is a two layer antiferromagnet with
intralayer exchange coupling $J_1$ and interlayer exchange coupling $J_2$.}
\label{figlay}
\end{figure}

In the past few years, there has been a significant interest in the
physics of quantum phase transitions in 2D spin
systems~\cite{Hertz,CHN,CSY,Alesha,Millis,CSS,IM}. The purpose of the
present communication is to study in detail the disordering
transition in
a two-layer $S=1/2$ Heisenberg antiferromagnet described by
\begin{equation}
H=  J_{1} \sum_{<i,j>, \alpha}\vec{S}_{\alpha,i}
\vec{S}_{\alpha,j} +J_{2} \sum_{i} \vec{S}_{1,i} \vec{S}_{2,i}
\quad . \label {H1}
\label{hamilt}
\end{equation}
Here $\alpha = 1,2$; the first sum runs over nearest-neighbors, and
the exchange couplings are assumed to be positive (see Fig.~\ref{figlay}).

For small $J_2/J_1$, the model describes two weakly interacting 2D
Heisenberg
antiferromagnets. Each of them is ordered at $T=0$ and possesses
Goldstone
excitations related to a spontaneous breakdown of a rotational
symmetry.
 In the opposite limit,
$J_2/J_1 \gg 1$, pairs of adjacent spins from different layers form
spin
singlets separated from triplet states by a gap, $\sim J_2$.
The presence of a gap implies that the rotational symmetry is not
broken.
Thus one should expect a disordering phase transition at some critical ratio
of $J_2/J_1$.

The two-layer Heisenberg model has attracted a lot of interest in the
last few
years~\cite{Sie,Tra89,Hid92,Mil93,San94,Mon95}
This interest was stimulated in part by the experimental
observation that some of the high-$T_c$ superconductors contain
pairs of
$CuO_2$ layers which are separated from other layers by a charge
reservoir~\cite{Sie,Tra89}. In addition, a two-layer
antiferromagnet is   probably the simplest
  $nonfrustrated$ spin system which
displays a quantum disordering transition of
 the $O(3)$ universality class. Several quantitative predictions
about the behavior of observables near such a transition have been made
recently~\cite{CSY},
and a two-layer antiferromagnet is an
ideal candidate to test these predictions.

The phase diagram of eq. (\ref{hamilt}) has been studied numerically,
by
Quantum Monte-Carlo~\cite{San94} and series expansion~\cite{Hid92}
techniques,
and analytically, using spin-wave~\cite{Tra89} and
mean-field Schwinger-boson
theory~\cite{Mil93}. There are several issues which emerged
from these studies. Some of them are related to the universal ratios
of
various observables and are discussed elsewhere~\cite{San95}.
Here we will
focus on the properties of the system at $T=0$, and on the
corrections to scaling at finite $T$. The key issue we want to address at $T=0$
is the applicability of  perturbative and self-consistent spin-wave
approaches
(the latter is very similar to the Schwinger-boson mean-field
theory). It is
well known that spin-wave expansion works extremely well for a
single-layer $S=1/2$ antiferromagnet.  At the same time,
for a two-layer system, spin-wave and Schwinger-boson theories yield
results which are inconsistent with numerical simulations. In
particular,
the Schwinger-boson mean-field theory yields
a critical value of the interlayer coupling
 $(J_2/J_1)_{cr} \approx 4.5$ ~\cite{Mil93}, which is
nearly two times larger than  $(J_2/J_1)_{cr}
\approx 2.5$ obtained in series expansion~\cite{Hid92} ($J^{cr}_2 =2.56$) and
Quantum Monte-Carlo~\cite{San94} ($J^{cr}_2 = 2.51$) studies.
 A self-consistent spin-wave
theory (see Sec~\ref{sw})
also predicts a very large value of  $(J_2/J_1)_{cr} \approx 4.3$.
The  spin-wave velocity at the
critical point,
$c_{sw} \approx 2J_1$ is also inconsistent with the Monte-Carlo data
which yield $c_{sw} \approx 1.7-1.8J_1$  (where we set the lattice
constant $a_0$ equal to unity).
We will argue that the
discrepancies between
the spin-wave results and the numerical simulations have a physical origin and
are related to
the fact that in the spin-wave approach one neglects longitudinal spin
fluctuations.
Our analytical approach to the problem
is based on the introduction of a triplet of
$S=1$ bosons for a pair of $S=1/2$ spins (see eq. (\ref{ML}) below).
In the disordered phase, this triplet of bosons describes the
excitations above the singlet ground state of a pair, while in the
ordered phase,
where we introduce a condensate for one type of boson,
the excitations are split into two transverse and one
longitudinal magnon modes. We will show  that the
contributions from longitudinal fluctuations to $J^{cr}_2$ and $c_{sw}$
are substantial, which
makes the $1/S$ expansion inapplicable. However, we  will also
show that as $J_2$ decreases, the spin-wave approximation becomes
more and more
reliable,
and at vanishing $J_2$, longitudinal spin fluctuations do not
contribute to the sublattice magnetization and susceptibility.

Furthermore, we will discuss the temperature dependence of the
uniform susceptibility, $\chi_u$, at the transition point.
Monte-Carlo simulations have shown that the universal, linear
behavior of $\chi_u$ at $J_2 = J^{cr}_2$
extends up to very high $T\sim J_1$. For comparison,
in a  single-layer antiferromagnet, the
deviations from linearity become substantial already at $T \sim
0.6J$~\cite{DM}. To understand this result, we will
compute the leading nonuniversal, $O(T^2)$ correction
to the susceptibility and show that it is numerically
quite small for all  $T < J_1$.

We start in
the next subsection with the spin-wave calculations for eq.
(\ref{hamilt}).
In  Sec~\ref{disord}, we will introduce the transformation to bosons
and consider in a systematic way the excitations in the disordered phase, the
critical value of $J_2$, and
the spin-wave velocity at the critical point.
In Sec~\ref{ord}, we
 extend the approach to the ordered state by introducing a condensate
for one of the bosonic fields.
We will show how the triplet of excitations
splits into two gapless transverse modes and a longitudinal mode with
a
finite gap. We will obtain the  $T=0$
sublattice magnetization and the uniform susceptibility at arbitrary $J_2$ and
show how they deviate from spin-wave results for increasing $J_2$.
 Finally, in Sec~\ref{qc},
we will compute the uniform susceptibility, $\chi_u (T)$ at
the critical point and show that the lattice-dependent,
$O(T^2)$ corrections to  the scaling form of
$\chi_u$ remain small up to $T= J_1$.
Our conclusions are stated in Sec~\ref{concl}.

\subsection{Spin-wave calculations}
\label{sw}

We start our considerations with a brief discussion
of the spin-wave calculations. At small $J_2$, the
spins are ordered antiferromagnetically  in the layers and  also
 between the layers.
Introducing bosons via the Holstein-Primakoff transformation,
and performing standard
manipulations, we obtain two branches of spin-wave excitations with the
dispersion $\epsilon_1 (k) = \epsilon_2 (k+\pi) = \epsilon_k$, where
 to order $1/S$,
\begin{equation}
\epsilon_k = 4{\bar J}_1 S[(1- \nu^2_k) + ({\bar J}_2/2
{\bar J}_1) (1 - \nu_k)]^{1/2} ,
\end{equation}
and $\nu_k = (\cos k_x + \cos k_y)/2$.
It is not difficult to show that the
fluctuations near $k=\pi$ are in-phase fluctuations of the spins in
the
two layers, while those near $k=0$ correspond to out-of-phase
fluctuations.
There is indeed a Goldstone mode in $\epsilon_k$
at $k=(\pi,\pi)$ because of a spontaneous
symmetry breaking.
The renormalized ${\bar J}_1$ and ${\bar J}_2$ differ
from the couplings in (\ref{hamilt}) due to the $1/S$ corrections:
\begin{equation}
{\bar J}_1 = J_1 \left( 1 - \frac{\delta_1 +
\delta_2}{S}\right);~~{\bar J}_2 =
J_2 \left( 1 - \frac{\delta_1 + \delta_3}{S}\right),
\label{3}
\end{equation}
 where
\begin{equation}
\delta_1 = \frac{1}{N} \sum_k~\frac{4{\bar J}_1S +{\bar
J}_2S}{2\epsilon_k} -
\frac{1}{2};~~\delta_2 = - \frac{1}{N} \sum_k \frac{(4 {\bar J}_1S
\nu_k + {\bar
J}_2S)\nu_k}{2\epsilon_k};~~\delta_3 = - \frac{1}{N} \sum_k
\frac{4 {\bar J}_1 S\nu_k + {\bar
J}_2S}{2\epsilon_k},
\label{4}
\end{equation}
The summation in (\ref{4}) is over the whole Brillouin zone.
The sublattice magnetization to order $1/S$ is given by $N_0 = S - \delta_1$.
 Evaluating $\delta_1$ with bare
couplings $J_{1,2}$, as it is required in the $1/S$ expansion,
we obtain that $\delta_1$ reaches a value of $S = 1/2$ only at a
very large $J_2/J_1 \approx 13.6$. A
somewhat better, though less justified estimate of $J^{cr}_{2}$ can
be obtained if one formally considers the expressions for the
renormalized couplings as
self-consistent equations, and solve them for $S=1/2$.
This procedure is
similar to the mean-field Schwinger-boson theory, and the
results we obtained are similar to those of Millis and
Monien~\cite{Mil93}: the sublattice magnetization first increases
with $J_2$,
passes through a maximum, and then decreases (see Fig.~\ref{figN_0}).
There is
a weak first-order disordering transition at $J^{cr}_2 \approx
4.36 J_1$. Still, the critical $J_2$ is much larger than
$J^{cr}_2 \approx 2.5 J_1$ obtained in numerical simulations.
We also computed the spin-wave velocity to $second$ order in $1/S$,
 and obtained after straightforward but somewhat tedious
calculations~\cite{denis}
\begin{equation}
c_{sw} = 2 \sqrt{2} S {\bar J}_1  \sqrt{1 + \frac{{\bar J}_2}{4{\bar
J}_1}} \left(1 + \frac{Q}{4S^2}\right),
\label{csw}
\end{equation}
where now ${\bar J}_{1,2}$ are the solutions of (\ref{3}) and (\ref{4})
to order $1/S^2$, and $Q$ is a cumbersome function of $J_2/J_1$
 whose explicit form we do not present.
 At $J_2 =0$, we
obtained $Q \approx 0.022$ which completely agrees with the results of other
studies~\cite{Iga92}.
The spin-wave velocity remains finite at the critical point, and it
is
therefore reasonable to compute it at
$(J_2/J_1)_{cr}$ suggested by numerical simulations.
Assuming for definiteness that $J^{cr}_2 = 2.55 J_1$,
 we obtained $Q \approx 0.044$. Evaluating then ${\bar J}_{1,2}$ and
substituting them
into (\ref{csw}), we find $c_{sw} \sim 3.62 J_1 S
(1 + 0.094/2S + 0.026/(2S)^2)$.
Observe that the $1/S^2$ correction is very small.
For $S=1/2$, we obtain $c_{sw} \sim 2.03~J_1$.
 As we mentioned earlier, this value is somewhat larger than
$c_{sw} \sim 1.7 - 1.8J_1$
extracted from the fit of the Monte-Carlo data for the uniform
susceptibility \cite{San94} to the scaling formula.

The main weakness of the spin-wave theory is that it assumes that
the long-range order is well established,
and only includes transverse spin fluctuations.
However, at the critical point, transverse and longitudinal
fluctuations
become indistinguishable and should be treated on equal ground. We
therefore proceed now
to perturbative calculations which explicitly take
 the longitudinal spin fluctuations into account.

\section{disordered phase}
\label{disord}

  The key starting point of our consideration is an observation that
for
sufficiently large $J_2$, pairs of adjacent
spins from the two planes form spin singlets. The excited state
of a given pair is a three-fold degenerate triplet state.
It is then natural to introduce a triplet of bosons for any given
pair. Each
boson describes the transformation from a singlet state to one of the
states with $S=1$. Specifically, we introduce
\begin{equation}
\vec{M_{i}}$=$\vec{S}_{1,i}+\vec{S}_{2,i},~~~~~~~
\vec{L}_{i}=\vec{S}_{1,i}-\vec{S}_{2,i},
\label{SML}
\end{equation}
 and three bosonic fields as
\begin{eqnarray}
M_{i}^{z} &=& a_{i}^{+}a_{i}-b_{i}^{+}b_{i}, ~~~~~~~~~~~~~~
L_{i}^{z} = - (c_{i}^{+}U_{i} +U_{i}c_{i}), \nonumber \\
M_{i}^{+} &=&\sqrt{2} (a_{i}^{+}c_{i}-c_{i}^{+}b_{i}), ~~~~~~~~
L_{i}^{+} = \sqrt{2} (a_{i}^{+}U_{i} +U_i b_{i}), \nonumber \\
M_{i}^{-} &=& \sqrt{2}(c_{i}^{+}a_{i}-b_{i}^{+}c_{i}), ~~~~~~~~
L_{i}^{-} =\sqrt{2} (b_{i}^{+}U_{i} +U_i a_{i}).
\label{ML}
\end{eqnarray}
where $U_i =\sqrt{1- a_{i}^{+}a_{i}
-b_{i}^{+}b_{i}-c_{i}^{+}c_{i}}$.
It is easy to check  that the
commutation relations  for  $\vec{M}$ and $\vec{L}$ are the same as
for a
vector and a generator of rotations: $[ M^{\alpha},M^{\beta}]=
i \epsilon_{\alpha  \beta
\gamma} M^{\gamma}; ~[ L^{\alpha},L^{\beta}] =i
\epsilon_{\alpha\beta\gamma}M^{\gamma}; ~[M^{\alpha},L^{\beta}] =i
\epsilon_{\alpha\beta\gamma}L^{\gamma}$. This in turn implies
that the spin commutation relations for $S_1$ and $S_2$ are
satisfied.
 The $U$
term, however, imposes the constraint that
 only one boson can be excited at each lattice site. This
indeed follows from the fact
that there are only four physical states
for a given pair of spins.
 For the physical states, we have $S^{2}_i = 3/4$ as
it should be. Notice that a similar restriction on the number of
bosons
holds also for the conventional
Holstein-Primakoff transformation for $S=1/2$. In this sense, the
transformation above can be
viewed as an extension of the Holstein-Primakoff transformation to
nonmagnetic states. One can
also introduce an analog to the Dyson-Maleev transformation,
but we found that the latter is less convenient for practical
purposes.

Furthermore,  a conventional way to perform spin-wave calculations for a
N\'{e}el-ordered state of a $S=1/2$ system is
 to extend a model to large $S$, perform $1/S$ expansion,
 and set $S=1/2$ at the very end of the calculations. We will now do
the same for a disordered state. To this end, we modify the
transformation to
bosons by
introducing a factor $\lambda \ll 1$ into the square root as $U_i  =
\sqrt{1 -
\lambda (a_{i}^{+}a_{i}+b_{i}^{+}b_{i}+c_{i}^{+}c_{i})}$, and
simultaneously introducing
an overall factor $1\sqrt{\lambda}$ into all three components of
${\vec L}_i$. It is not difficult to check that the commutation
relations
between ${\vec L}$ and ${\vec M}$ (and, hence, the spin algebra) do
not change
under this transformation; however, the value of the spin on each site
in the
ground state is now $O(1/\lambda) \gg 1$.
Below, we perform a systematic perturbative expansion in $\lambda$.
 The physical results, however, correspond only to $\lambda =1$.

Eq. (\ref{ML}) has been applied
before to study the dimerization in the $S=1/2$ Heisenberg model
on a square lattice with an
interaction between first and second neighbors~\cite{Chu91},
 and also the dimerization
transition in a $S=1$ chain~\cite{Chu91b}.
 We believe that this approach
 has some advantages over the
mean-field Schwinger-boson theory. For example,
 it correctly reproduces the fact that at the critical point and in
the
disordered phase, the magnon excitation spectrum is three-fold
degenerate.

We now substitute (\ref{SML})
and (\ref{ML}) into the Hamiltonian. To leading order in
${\lambda}$, the interaction between bosons can be neglected,
 and diagonalizing the quadratic form in bosons, we
obtain a
three-fold degenerate excitation spectrum with the dispersion
\begin{equation}
\epsilon_k = \sqrt{A^2_k - B^2_k},
\end{equation}
 where $A_k = J_2 + 2 J^*_1 \nu_k$, and $B_k =
2 J^*_1 \nu_k$, and $J^*_1 = J_1/\lambda$. For sufficiently large
$J_2$,
the excitation energy is
real (which indicates a stability), and there is a finite gap in the
spectrum
whose minimum is at $k = \pi$. This gap vanishes
 at $J_2 = J^{cr}_2 = 4 J^*_1$. Below this point, the excitations
near $k =
\pi$ are purely imaginary which
signals an instability and implies a need for a change of the ground
state.

To obtain a better estimate for the critical value of $J_2$, we
included anharmonic terms into consideration, computed
the self-energy terms by usual
means, and obtained to order $O(\lambda^2)$:
\begin{equation}
J^{cr}_2 = 4 J^*_1~ \left( 1 - 0.665\lambda +
\frac{1}{\pi^2}~\lambda^2
\log{1/\lambda} + O(\lambda^2) \right)
\label{eq21}
\end{equation}
We see that the first-order correction shifts the transition towards
smaller $J_2$. If we had restricted the calculation to include only this term,
we would obtain $J_2^{cr}$ in a range
between $1.34 J_1$ and $2.4J_1$, depending on whether we leave the
correction
in the numerator or put it into the denominator.
The second-order correction is
{\it positive} and
partly compensates the downshift renormalization due to the
first-order
term. Unfortunately, the second-order correction is logarithmically
divergent at the transition~\cite{co}, and we
cannot obtain  the precise value  of $J^{cr}_2$ to order
$\lambda^2$. We therefore can only argue that
the actual value of $J^{cr}_2$ is in
between our zero-order and first-order results.  A somewhat better
estimate of $J^{cr}_2$ can be obtained approaching the transition from the
ordered phase, and will be discussed in appendix A. Notice, however, that
the first-order estimate of $J^{cr}_2$ is already closer to the numerical
result than $J^{cr}_2 \sim 4.3 J_1$
 which was obtained in a self-consistent spin-wave approach.

We also
computed the spin-wave velocity at the critical point. To order
$O(\lambda)$, we obtained
\begin{equation}
c_{sw} = 2~J^{*}_1  \left ( 1 - 0.256~ \lambda \right)
\label{eq3}
\end{equation}
For the physical case of $\lambda =1$, this gives $c_{sw}$
 between $1.49J_1$
and $1.59J_1$ again depending on whether we keep the
 correction in the numerator or
put it into the denominator.
The second order correction  to the spin-wave velocity is again
positive
and partly compensates the $O(\lambda)$ contribution, but it is
again of the form $\lambda^2 \log{1/\lambda}$ which prevents us from
obtaining the precise value of $c_{sw}$ to order $O(\lambda^2)$.
Alternatively however, we can reexpress $c_{sw}$
 in terms of the critical value of $J_2$. Doing this, we find that to
order
$\lambda$, $c_{sw} =
0.5 J^{cr}_2  (1 + 0.409 \lambda + \ldots)$. For $\lambda =1$, this
yields
$c_{sw} =  0.705 J^{cr}_2 $.
Using then the numerical result $J^{cr}_2 = 2.55 J_1$,
we obtain
$c_{sw} \sim 1.80 J_1 $ which is consistent with
 $c_{sw}= 1.7-1.8 J_1 $, extracted from the Monte-Carlo data.
In any case, the velocity we found is smaller
than that obtained in the spin-wave theory.

\section {ordered phase}
\label{ord}

We now consider the case $J_2 < J^{cr}_2$ when the system
possesses
a N\'{e}el order. We assume that the sublattice magnetization, $N_0$, is
directed along the $z$-axis. In our approach, a nonzero $N_0 \equiv N^{z}_0$
implies that there is a
single-particle condensate of the $c$-quanta with momentum
$\pi \equiv (\pi,\pi)$: $<c_{\pi}> = \alpha$.
 In a mean-field
approximation, we then have $N_0 = \lambda^{-1}  \sqrt{\beta (1 -
\beta)}$, where
$\beta = \lambda \alpha^2$. Introducing the condensate into the
Hamiltonian and evaluating the ground state energy, $E_0$,
 in the mean-field approximation
(i.e., to  leading order in $\lambda$, but keeping
$\beta$ fixed), we find
\begin{equation}
\lambda E_0 = J_2 \beta - 4 J^*_1 \beta (1 - \beta)
\end{equation}
Minimizing the energy, we obtain
$\beta = \beta_0 = (4 J^*_1 - J_2)/8J^*_1$.
For $J_2=0$, we  have $\beta_0 = 1/2$,
and hence $N_0 = 1/(2 \lambda)$ as it should be.
We then performed the
 standard computations for a Bose gas with a
condensate and
 obtained the quasiparticle spectrum.
It now contains two different branches of quasiparticle excitations.
The excitation spectrum for fluctuations in the direction
perpendicular to the condensate (i.e., for $a-$ and $b-$type bosons) is doubly
degenerate.
For these excitations, we obtained to leading order in $\lambda$
\begin{equation}
\epsilon_{\perp} (k) = 4 J^*_1 ~(1 - \beta)~\left[(1 + \nu_k)~
 (1- \frac{\beta}{1-\beta}~\nu_k)\right]^{1/2}
\label{perpdisp}
\end{equation}
We see that the transverse fluctuations are gapless as they indeed
should be.
For the spin-wave velocity near $k=\pi$ we have
\begin{equation}
c_{sw} = 2J^*_1 \left(1 - \beta \right)^{1/2}
\label{cswmf}
\end{equation}
Observe that for $J_2 =0$ we recover the  mean-field dispersion for the
Heisenberg
antiferromagnet: $\epsilon_{\perp} = 2 J^*_1 (1 - \nu^2_k)^{1/2}$.

For the dispersion relation of
the fluctuations along the
direction of the condensate (i.e., for $c$-type bosons), we found
\begin{equation}
\epsilon_{\parallel} (k) = 4 J^*_1~\left( 1 + (1 -2 \beta)^2 \nu_k
\right)^{1/2}
\label{pardisp}
\end{equation}
We see that the longitudinal fluctuations
in the ordered phase have a finite gap at the antiferromagnetic
momentum,
 $\epsilon_{\parallel} (\pi) = 8J^*_1~[\beta~(1 - \beta)]^{1/2}$.
Also observe that at $J_2 =0$,
the longitudinal mode becomes  dispersionless: $\epsilon_{\parallel} (k)
= 4 J^*_1$. However, we do not know whether this result survives beyond the
leading order in $\lambda$. The actual dispersion for a $c$-boson
may also contain some finite imaginary part (due to higher-order
terms in
$\lambda$) which can be substantial at small $J_2$.

The computations which lead to  eq. (\ref{pardisp})
require some care. The important point is that
since $\alpha \sim \lambda^{-1/2}$, there is a cancellation of the
overall factor
$\lambda^n$ in the n-th term in the expansion over density in $U$,
 and all terms  in the
series  are in fact relevant. In practice, this implies that
evaluating the contribution
to the longitudinal dispersion from $L_z L_z$, one has to examine
each term in
the series, put all $c-$bosons except for two  into a
condensate,
compute the numerical combinatoric factor, and explicitly sum  the resulting
series.

We then used the results for the quasiparticle spectra and
 computed the sublattice magnetization and the uniform spin
susceptibility beyond the mean-field level, to
order
$O(\lambda)$. The computations and
the procedure of extending the first-order
results
to $\lambda =1$ are discussed at some length
in the appendix. The results
are presented in Fig.~\ref{figN_0} and Fig.~\ref{figchi}. For comparison,
in Fig.~\ref{figN_0},
we also plotted
the self-consistent spin-wave result for the magnetization. It is
essential that at $J_2 =0$, both our results
are $exactly$ the same as  obtained in the first-order $1/S$ expansion.
In other words, for a single layer antiferromagnet,
there are no independent contributions from
longitudinal fluctuations.  This result provides a qualitative explanation of
why the $1/S$ expansion
works so well for a single layer antiferromagnet. Indeed,
in our approach, we treat longitudinal fluctuation as a separate bosonic mode.
At the same time, in the $1/S$ expansion, the longitudinal mode
appears as  a pole in the two-particle Green function. To obtain
this pole, one has to sum an infinite number of the $1/S$ terms.
Then, roughly speaking, the contribution from the longitudinal mode represents
the contributions from high-order terms in the $1/S$ expansion.
The absence of the longitudinal correction in our effective
``spin-wave theory" therefore implies that
the series in $1/S$ converges rapidly, and
the dominant contribution comes from the first-order term.

We emphasize however that the longitudinal fluctuations can be neglected only
for $J_2/J_1 \ll 1$. As
$J_2$ increases, the deviation of our result for
$N_0$ from the self-consistent spin-wave result
 becomes more and more substantial as seen in
Fig.~\ref{figN_0}. Near the disordering transition, longitudinal and
transverse fluctuations have nearly equal strength, and the
actual behavior of sublattice magnetization and uniform susceptibility
differs in an essential way from the
prediction based on the spin-wave  theory.

Notice that in some range of small $J_2$,
both the sublattice magnetization
and the uniform susceptibility are larger than for a single layer, i.e., the
system first becomes more ``classical'', and only then, at larger $J_2$,
do quantum fluctuations push the system towards the disordering transition.
The region of more  ``classical'' behavior at intermediate $J_2$
has been observed in the mean-field Schwinger-boson approach~\cite{Mil93};
it is also present in the self-consistent spin-wave analysis (see
Fig.~\ref{figN_0}).

Near the transition point, we obtained
\begin{equation}
N_0 = \frac{Z_{N}}{\lambda}~\sqrt{\beta},~~~~\chi_{\perp} = A
(Z_{\chi}/4J_1)~(\beta/\lambda^2)^{1/(1 +\eta)},
\label{NN}
\end{equation}
where $Z_{N} = 1 - 0.163 \lambda$,  $Z_{\chi} = 1 + 0.255\lambda$,
and $\eta \approx 0.03$ is the critical exponent for spin
correlations at criticality. The factor $A$ cannot be obtained within
the present approach because of the divergence of the Gaussian corrections
near the transition point in $2+1$ dimensions. Our estimates in the appendix
place $A$ to be roughly equal to $2$. The  ratio
$N^2_0/[2\pi (\rho_s)^{1/(1+\eta)}]$ is an overall factor for the
dynamical spin susceptibility. Using (\ref{NN}) and the result for
the spin-wave velocity at the
transition point, we obtain
$N^2_0/[2 \pi (\rho_s)^{1+\eta}] = B/J_1^{1+ \eta}$, where
$B = (1 - 0.06\lambda)/(2 \pi A^{1+ \eta})$.
Three different numerical estimates of $B$ all yield $B = 0.063$~\cite{San95}.
This is roughly consistent with our estimate $B = 0.149/A$, though we
only approximately know that $A \sim 2$.

We also computed the quasiparticle dispersion to order $O(\lambda)$
near
the transition, and explicitly obtained the Goldstone mode in the
transverse channel. These calculations were performed only to leading
order in
$\beta$, when one can neglect
cubic terms. For a general $\beta$,
the Goldstone modes arise as a result of cancellations between the
the second-order contributions from
the cubic terms
and the first-order contributions from the quartic terms.
A similar situation is known to exist in frustrated spin
systems~\cite{frustr}.
We did not perform explicit calculations of the spin-wave spectrum at
arbitrary
$\beta$ and therefore cannot make a definite prediction about how
longitudinal
fluctuations influence the spin-wave velocity at small $J_2$.
However, given the good agreement between our result and the spin-wave result
for the susceptibility in a  single
layer antiferromagnet, and the consistency between the spin-wave
result for
the spin stiffness, $\rho_s = c^2_{sw} \chi_{\perp}$, and the
numerical data~\cite{singh},
we expect the corrections due to
longitudinal fluctuations to be zero or at least small at vanishing
$J_2$.
However, near the transition point, we have already shown
that the  corrections to the spin-wave velocity
cannot be reduced to only those due to transverse fluctuations. Thus the
spin-wave result for $c_{sw}$, which neglects longitudinal
contributions, is most probably not quite accurate even
 though the velocity remains finite at the transition point,
and the $O(1/S^2)$ correction to $c_{sw}$ is  much smaller than the
$O(1/S)$ correction (see Sec.~\ref{sw}).  In other words, we argue
that
near the transition, the series of $1/S$ terms is $not$ rapidly
convergent even
if the first few terms in the series seem to indicate the contrary.

\section {uniform susceptibility at the critical point}
\label{qc}

In a single-layer Heisenberg antiferromagnet, the linear temperature
dependence of
the uniform susceptibility associated with quantum-critical behavior
 has been observed in the temperature range $0.35 J_1 < T <
0.6J_1$. At lower temperatures, there is a crossover to another
linear
 behavior associated with the renormalized-classical regime (which,
however,
has not yet been observed),
while at higher temperatures, $\chi_u$ flattens and has a broad
maximum at
$T \sim J_1$~\cite{DM,Johnston}.
How far  the linear dependence extends at high $T$
depends on the lattice-dependent corrections to
scaling. The  Monte-Carlo results for
a two-layer antiferromagnet at the critical
$J_2$ have shown that
the linearity extends to  sufficiently high temperatures, $T \sim
J_1$, i.e., the corrections to scaling at $J_2 = J^{cr}_2$
are smaller than those of a single-layer antiferromagnet. Below we will
compute these corrections perturbatively. But first we consider the
sigma-model description of a two-layer system, from which
one can obtain the leading, universal, linear in $T$ term in the uniform
susceptibility.

\subsection{Sigma-model analysis}

A simple way to obtain a sigma-model description of a spin-$S$
quantum antiferromagnet, which we will follow,
 was  suggested by Affleck~\cite{Aff86}.
In application to  our system, one has
to double a unit cell in each of the two layers and
introduce ${\vec n}_{\alpha,i} =
({\vec S}_{\alpha,i} - {\vec S}_{\alpha,i+1})/2S, ~{\vec
l}_{\alpha,i} =
({\vec S}_{\alpha,i} + {\vec S}_{\alpha,i+1})/2S$. At large $S$,
${\vec n}$
becomes a classical unit field with commuting components, while
the commutation relations
between $\vec n$ and $\vec l$ are the same as for a vector and a
generator of
rotations. Introducing ${\vec n}_{\alpha}$ and ${\vec l}_{\alpha}$
 into the Heisenberg Hamiltonian and  making a transformation from the
 Hamiltonian to the corresponding action which contains only
the derivatives of ${\vec n}_{\alpha}$, we  obtain
the action of two interacting $O(3)$ sigma-models.
In terms of $\vec n$ and $\vec l$, the interaction term has the
form $\Xi^2 ({\vec n}_1 {\vec n}_2 -{\vec l}_1 {\vec l}_2)$, where
$\Xi^2
\propto J_2$.  The generator of rotations itself
contains a derivative of $\vec n$ :
 ${\vec l} \sim {\vec n} \times \frac{\partial {\vec n}}{\partial
\tau}$, and
the ${\vec l}_1 {\vec l}_2$ term thus only leads to a velocity
renormalization.
Neglecting this term, and also introducing the magnetic field into
the action for susceptibility calculations, we obtain
\begin{equation}
{\cal S} = \frac{1}{2 g} \left[ (\nabla {\vec n}_1)^2 + (\nabla {\vec
n}_2)^2 +
\Xi^2 {\vec n}_1 {\vec n}_2 + \frac{1}{c^2_{sw}}
\left( \frac{\partial {\vec n}_1}{\partial \tau} - i {\vec H} \times
{\vec n}_1\right)^2 + \frac{1}{c^2_{sw}}
\left( \frac{\partial {\vec n}_2}{\partial \tau} - i   {\vec H}
\times
{\vec n}_2\right)^2 \right]
\label{nf}
\end{equation}
where
$g$ is a coupling constant which depends on the ratio $J_2/J_1$, and
$H$ is measured in units of $g \mu_B /\hbar$.
Introducing ${\vec \sigma}_{1,2} = ({\vec n}_1 \pm
 {\vec n}_2)/\sqrt{2}$,
we can rewrite the sigma-model action as
\begin{equation}
{\cal S} = \frac{1}{2 g} \left[ (\nabla {\vec \sigma}_1)^2 +
(\nabla {\vec \sigma}_2)^2 +
\Xi^2 {\vec \sigma}^2_1 + \frac{1}{c^2_{sw}}
\left( \frac{\partial {\vec \sigma}_1}{\partial \tau} - i {\vec H}
\times
{\vec \sigma}_1\right)^2 + \frac{1}{c^2_{sw}}
\left( \frac{\partial {\vec \sigma}_2}{\partial \tau} - i {\vec H}
\times
{\vec \sigma}_2\right)^2 \right]
\label{sigma}
\end{equation}
 The constraints on the $\sigma$-fields are
$\vec {\sigma}_1 \vec {\sigma}_2 =0, ~ {\vec \sigma}^2_1 +
{\vec \sigma}^2_2 = 2$.

The evaluation of the susceptibility at the mean-field ($N = \infty$) level
is straightforward. Using the results of~\cite{CSY},
we obtain $\chi_u = (\chi_1 + \chi_2)/2$, where
$\chi_u$ is a susceptibility {\it per spin}, and
$\chi_{1,2}$ are the mean-field susceptibilities for the two
sigma-fields
\begin{equation}
\chi_{1,2} = \frac {T}{\pi c^2_{sw}} \left[ \frac{c_{sw} m_{1,2}}{T}~
\frac{e^{c_{sw}
m_{1,2}/T}}{e^{c_{sw} m_{1,2}/T} -1} - \log\left(e^{c_{sw} m_{1,2}/T}
-1\right)\right]
\label{chim-f}
\end{equation}
where $m_{1} = \sqrt{\Xi^2 + m^2}$, and $m_2 =m$, where $m$ is the
mass
obtained from the second constraint equation. At $g = g_c = 8 \pi
(\Lambda +
\sqrt{\Lambda^2 + \Xi^2} - |\Xi|)^{-1}$ where $\Lambda \sim J$ is the
upper cutoff, we have
$m = \Theta T + O(T^2)$,
 where~\cite{SY} $\Theta = 2 \log [(\sqrt{5} +1)/2]$. At low $T \ll
\Xi$, $\chi_1$ is
exponentially small in $T$ and can be neglected compared to $\chi_2$.
It is
not difficult to show that the
contributions related to the fluctuations of $\sigma_1$ are exponentially small
and  persist  even
 beyond the mean-field level. As a result, the universal term in the
uniform susceptibility is solely due to $\sigma_2$, and $\chi_u$
is precisely $half$ of that in a single-layer model.

\subsection{Computation of the subleading term in $\chi_u (T)$}

The sigma-model approach gives us the leading, universal,
temperature dependence of the uniform susceptibility. Now we
compute the leading nonuniversal correction to $\chi_u$. We will again use a
microscopic approach based on a transformation to bosons.
However, this approach clearly
has to be modified compared to what we did before at $T=0$
 because the quasiparticle
 densities (both normal and anomalous) diverge at finite
temperature, and
the expansion in $\lambda$ is no longer valid. For this reason,
we will perform
a self-consistent, mean-field calculation of the susceptibility:
we first assume that anharmonic
contributions to the quasiparticle spectrum produce a  $T-$dependent
gap which eliminates  divergencies of quasiparticle densities
at the transition point, then we
evaluate the quasiparticle densities
with the renormalized spectrum, and
solve the self-consistent equations for the
gap. In principle, one can perform these calculations using the
same transformation to bosons as before. This procedure is then
equivalent to self-consistent ``$1/S$'' calculations in 2D~\cite{Takah}.
However, we found it more convenient to use
a similar but slightly different form of the transformation to
bosons,
 introduced by Bhatt and Sachdev~\cite{Bha91}.
In their approach, one introduces an
extra bosonic field instead of a $U-$ term in (\ref{ML}):
\begin{equation}
L_{i}^{z} =- (c_{i}^{+}s_{i} +s_{i}^{+}c_{i});~~~
L_{i}^{+} =\sqrt{2} (a_{i}^{+}s_{i} + s_{i}^{+} b_{i});~~~
L_{i}^{-} =\sqrt{2} (b_{i}^{+} s_i + s_{i}^{+} a_{i}).
\end{equation}
 The expressions for
${\vec M}$ are the same as before. The commutation algebra for spins
is again
satisfied, while the constraint on the length of the spin now reduces
to
$a_{i}^{+}a_{i} + b_{i}^{+}b_{i} + c_{i}^{+}c_{i} + s_{i}^{+}s_{i}
=1$.
The advantage of this transformation is that one no longer needs to
assume that the density of excitations is small.
However, we did not use this transformation for our $T=0$ calculations above
because we found it
difficult to perform a systematic expansion about the mean-field
solution.  However, the mean-field calculation is straightforward:
one has to put the $s-$field into a condensate ($<s> = s_0$),
neglect fluctuations of $s$, and
 reduce the on-site constraint to a constraint imposed on
average quantities. We first list the $T=0$ results which are
 similar (but not identical)  to the results we
obtained to the zeroth order in $\lambda$. In the disordered phase, we indeed
again find
the three-fold degenerate
quasiparticle spectrum with  $\epsilon_k =
\sqrt{A^2_k - B^2_k}$, where $A_k = J_2 + 2 J_1 s^2_0 \nu_k$,
$B_k = 2 J_1 s^2_0 \nu_k$, and
the self-consistent equation for $s_0$ follows from the constraint on the
length of the spin:
$s^2_0 = 1 - (3/N) \sum_k (A_k - \epsilon_k)/2 \epsilon_k$.
At the transition point, we obtained
$s_0 \approx 0.9$. The critical value of $J_2$ is then
 $J^{cr}_2 = 4 J_1 s^2_0 \approx 3.2 J_1$, and the $T=0$
spin-wave velocity at criticality is $c_{sw} = 2J_1 s_0 \approx
1.8J_1$.

We now consider finite temperatures. Assume
that the condensate of the $s-$field has a form
 $s^2_0 = (s^2_0)_{T=0} ~(1 - m^2/4)$, such that
at the critical point and near
 $k = \pi$, $\epsilon^2_k = c^2_{sw} (k^2 + m^2)$. Substituting the
full
expressions for $A_k$ and $\epsilon_k$ into a self-consistency
equation at
finite $T$, expanding in $T$ and evaluating the lattice sums, we
obtain
\begin{equation}
\frac{c_{sw} m}{T} = \Theta \left( 1 + \mu~\frac{T}{J_1} + O(T^2)
\right)
\end{equation}
Here $\Theta$ is the same
as in the sigma-model calculations, and
the second term is a lattice-dependent correction
which we found to be $\mu = -0.061$.
Furthermore, we have checked that the mean-field formula for the uniform
susceptibility is given precisely by eq. (\ref{chim-f}) with no extra
lattice-dependent corrections (we
applied a magnetic field,
rediagonalized the quadratic form in bosons,
and computed the magnetization along the
field). Substituting then the result for the mass $m$ to order $T^2$
into (\ref{chim-f}), we obtained
\begin{equation}
\chi_u = Q~\frac{T}{c_{sw}^2} \left ( 1 -
\left(\frac{2\Theta \mu}{\sqrt{5}}\right) ~\frac{T}{J_1} +
 O(T^2) \right),
\end{equation}
 where $Q = \sqrt{5} \Theta /4\pi$ in the mean-field approximation
(the $1/N$ correction extended to a physical case of $N=3$
reduces this value by about 20\%~\cite{CS}).
 We see that the
numerical factor in the subleading term in the susceptibility is
very small, and, e.g., at $T =
J_1$, constitutes only 5\% of the mean-field value.
Indeed, at  $T\sim J_1$, higher-order corrections
in $T/J_1$ could also be relevant,  but the fact that the leading
correction to the scaling result is small is at least an indication that
the  universal linear dependence of the uniform susceptibility
 extends to sufficiently high $T\sim J_1$. As we already
discussed, this
is consistent with the
 Monte-Carlo data~\cite{San94}.

\section{conclusions}
\label{concl}

In this paper, we considered a two-layer Heisenberg
antiferromagnet which can either be  in the N\'{e}el-ordered or in the
disordered
phase at $T=0$ depending on the ratio of the intra- and interlayer
exchange constants.
 We applied a transformation to bosons which
is suitable for a singlet configuration of a pair of spins, and
considered in
a systematic expansion the quasiparticle excitations in the
disordered phase,
and the critical value of the interlayer coupling.
We then extended the approach to the ordered phase by introducing a
single-particle condensate of one of the Bose fields and computed
the mean-field quasiparticle dispersion, the sublattice magnetization and
 the transverse susceptibility  at arbitrary $J_2$. We  then computed one-loop
corrections to the sublattice magnetization and the susceptibility, and
considered the relative strength of the longitudinal spin fluctuations.
 We found that  the contributions of these fluctuations
 are zero in  a single-layer
antiferromagnet, but are quite substantial  near the transition
point, where  the transverse and the longitudinal  fluctuations are
equally important.
The results of our $T=0$ calculations are in a reasonable agreement with
the Monte-Carlo and series expansion data.
 We also computed the temperature dependence of the
uniform susceptibility at the critical point,
and  found that the lattice-dependent corrections to the universal
scaling
behavior $\chi_u \propto T$ are small for all $T \leq J_1$. This is
again
consistent with the Monte-Carlo data which show that
 the linear behavior of $\chi_u$ extends to sufficiently high
temperatures $T\sim J_1$ and flattens only at even higher temperatures.

It is our pleasure to thank A. Millis, H. Monien, S. Sachdev, A.
Sandvik and A. Sokol for useful conversations.

\section{appendix}
In this appendix, we compute the sublattice magnetization, the
transverse
susceptibility and the spin stiffness in the
N\'{e}el phase to order $\lambda$. We start with the calculations of the
magnetization.

\subsection{Sublattice magnetization}

Our point of departure is the expression for $L_z$, eq. (\ref{ML}),
extended to $\lambda \ll 1$. In the ordered phase,
$N_0 = <L_z>/2 = <c^+U>/\sqrt{\lambda}$, where the averaging is over
the exact ground state.
The mean-field calculations in the ordered state were
 presented in Sec.~\ref{ord}. In these calculations,
 we considered only the  condensate piece of the $c-$field,
$ <c> = \alpha$. Here we
will need both, $\alpha$
and the fluctuating component of $c$. Substituting
$c_k = \alpha \delta_{k,\pi}
+ {\tilde c}_k$, into $N_0$,
expanding in $U$ up to an infinite
order, and collecting all terms which contain
 at most one pair product of fluctuating
fields, we obtain after some simple combinatorics
\begin{equation}
N_0 = \frac{\sqrt{\beta (1 - \beta)}}{\lambda}~\left[1 - \lambda
\left( Z_1 (\beta) + Z_2 (\beta) +
  Z_3 (\beta) + Z_4 (\beta) \right)\right]
\label{N_00}
\end{equation}
where, we recall, $\beta = \lambda \alpha^2$, and
\begin{eqnarray}
Z_1(\beta) &=& \frac{\beta}{8(1 - \beta)^2}  \nonumber \\
Z_2 (\beta) &=& -\frac{2 - \beta}{4(1 - \beta)^{2}}~ \frac{1}{N}
\sum_k
\frac{B_{\parallel} (k)}{2 \epsilon_{\parallel} (k)} \nonumber \\
Z_3 (\beta) &=& \frac{4 - 3\beta}{4(1 - \beta)^{2}}~
\frac{1}{N} \sum_k
\left(-\frac{1}{2} + \frac{A_{\parallel} (k)}{2 \epsilon_{\parallel}
(k)}
\right) \nonumber \\
Z_4 (\beta) &=& \frac{1}{1 - \beta}~\frac{1}{N} \sum_k
\left(-\frac{1}{2} + \frac{A_{\perp} (k)}{2 \epsilon_{\perp} (k)}
\right).
\label{Z}
\end{eqnarray}
Here
\begin{eqnarray}
A_{\parallel} (k) &=& J_2 + \frac{2 J^*_1}{1- \beta}~\left(\beta (4 -
3
\beta) + \nu_k (1 -2 \beta)^2 \right), \nonumber \\
B_{\parallel} (k) &=& \frac{2 J^*_1}{1- \beta}~\left(\beta (2 -
\beta) + \nu_k (1 -2 \beta)^2 \right), \nonumber \\
A_{\perp} (k) &=& J_2 + 2 J^*_1 \nu_k
+ 4J^*_1 \beta (1 - \nu_k),
\label{ZZ}
\end{eqnarray}
$J^*_1 = J_1/\lambda$,
and the dispersions for transverse and longitudinal fluctuations are
given by
(\ref{perpdisp}) and (\ref{pardisp}).
Observe that near the critical point, $N_0 \propto \sqrt{\beta}$.
The next step is to express $\beta$ in terms of $J_1$ and $J_2$.
 To this end, we compute the
ground state energy, $E_0$ with the $O(\lambda)$
corrections which come from noninteracting spin-waves {\it and} from
the normal ordering of $c-$operators in the expansion of $U$.
Combining the two contributions, we obtain
\begin{eqnarray}
\lambda E_0 &=& J_2  \beta  - 4J^*_1 \beta [(1 - \beta) - 2 \lambda
(1- \beta)~Z_1 (\beta)] \nonumber \\
&& - \lambda~\sum_k [A_{\perp} (k) -
\epsilon_{\perp} (k)] - \frac{\lambda}{2}~\sum_k [A_{\parallel} (k) -
\epsilon_{\parallel} (k)]
\end{eqnarray}
Minimization with respect to $\beta$ then yields
\begin{equation}
\beta = \beta_0 - \lambda \left(Z_5 (\beta_0) + Z_6 (\beta_0) + Z_7
(\beta_0) + Z_8 (\beta_0) \right)
\label{beta}
\end{equation}
where
\begin{eqnarray}
Z_5 (\beta_0) &=& \frac{1}{N} \sum_k ~\left (1 - \nu_k \right)
\left(-\frac{1}{2} + \frac{A_{\perp} (k)}{2 \epsilon_{\perp} (k)}
 \right) \nonumber \\
Z_6 (\beta_0) &=&
\frac{1}{4 N} \frac{1}{(1-\beta)^2}\sum_k ~\left(  4-6\beta+3 \beta^2
+(-3+8\beta-4\beta^2)\nu_k \right)
\left(-\frac{1}{2} +
\frac{A_{\parallel} (k)}{2 \epsilon_{\parallel} (k)}\right) \nonumber
\\
Z_7 (\beta_0) &=& - \frac{1}{4 N} \frac{1}{(1-\beta)^2}\sum_k
{}~\left(2-2\beta+\beta^2 +(-3+8\beta-4\beta^2)\nu_k  \right)
\frac{B_{\parallel} (k)}{2 \epsilon_{\parallel} (k)} \nonumber \\
Z_8 (\beta_0) &=&  \frac{\beta(2- \beta)}{8~(1 -
\beta_0)^2}~
\end{eqnarray}
Notice that the correction terms
 $Z_{1}, Z_2, Z_{3}, Z_{6}, Z_7$  and $Z_8$ are due to
fluctuations in the direction of the condensate, while the terms
$Z_4$ and $Z_5$ come from transverse fluctuations.

Substituting (\ref{beta}) into (\ref{N_00}), we obtain, to order $O(\lambda)$
\begin{equation}
N_0 = \frac{\sqrt{\beta_0 (1 - \beta_0)}}{\lambda}~\left[\frac{1 -
\lambda~Z_b/\beta_0}{1 - \lambda~Z_b/(1 - \beta_0)}\right]^{1/2}~\left(1 -
\lambda Z_a\right)
\label{N_0}
\end{equation}
where
\begin{eqnarray}
Z_a &=& Z_1 (\beta_0) + Z_2 (\beta_0) +  Z_3 (\beta_0)+ Z_4 (\beta_0)
\\
Z_b &=& Z_5 (\beta_0) + Z_6 (\beta_0) + Z_7(\beta_0) + Z_8 (\beta_0)\nonumber
\end{eqnarray}

At $J_2 =0$, $\beta_0 = 1/2$, and evaluating the lattice sums, we obtain
$N_0 = (1/2\lambda) - n_0$, where $n_0 = N^{-1} \sum_k ((1 - \nu^2_k)^{-1/2}
-1)/2 = 0.197$ is
the density of transverse fluctuations (spin waves)~\cite{Iga92}.
This result is equivalent to the first order
spin-wave result, i.e., longitudinal fluctuations
do not contribute to sublattice magnetization to first order in $\lambda$.
This is a  direct consequence of
the fact that the longitudinal mode is dispersionless at $J_2 =0$,
and hence
the $c-$bosons on adjacent sites do not interact with each other.
It is essential, however, that the longitudinal fluctuations are small
only for $J_2/J_1 \ll 1$.
Near the disordering transition, longitudinal and
transverse fluctuations have nearly equal strength, and the
actual behavior of magnetization differs in an essential way from the
prediction based on the spin-wave  theory.
In this limit, we obtained
\begin{equation}
 N_0 = \frac{Z_{N}}{\lambda}~\sqrt{\beta},
\label{N}
\end{equation}
where $Z_{N} = 1 - 0.163 \lambda$, and
the fully renormalized $\beta$
satisfies the equation
\begin{equation}
8J_1\beta (1 -(3/\pi)~\lambda/\sqrt{\beta_0}
+ ...) = J^{cr}_2 - J_2 ,
\label{aa}
\end{equation}
 where $J^{cr}_2 =
4 J^*_1 ( 1 - 0.665 \lambda + \ldots)$ is the same as we obtained
approaching
the critical point from the disordered phase.
 We have checked that
the  two
analytical expressions for $J^{cr}_2$ are indeed also identical. The
subleading term in (\ref{aa})
is a Gaussian correction to the sublattice
magnetization. In the theory of phase transitions, it is usually assumed
that the Gaussian term is in fact expressed in terms of fully renormalized
$\beta$ rather than $\beta_0$.
The correction term then diverges as one approaches $J^{cr}_2$
as it indeed should in 2+1 dimensions. Due to this divergence, the
 self-consistent approach is
valid only at $J^{cr}_2 - J_2 > \lambda^2$.
In the opposite limit $J^{cr}_2 -J_2 \ll \lambda^2$, scaling considerations
predict that the
sublattice magnetization should behave as
$N_0 \sim (J^{cr}_2 - J_2)^{{\bar \beta}}$, where ${\bar \beta} \sim
0.35$.

The above considerations are also relevant
as to how one should extend the
perturbative result for $N_0$ at arbitrary $\beta$ to $\lambda =1$.
We have seen that
near $J_2 =0$, one should keep $\beta = \beta_0$
in the $O(\lambda)$ terms. At the same time, it is not difficult
to make sure that in order
to obtain the same $J^{cr}_2$ on both sides of the transition,
 one has to perform calculations self-consistently, i.e.,
evaluate the subleading terms in (\ref{beta})
with the
fully renormalized $\beta$. To first order in $\lambda$, both procedures
are indeed equivalent. However, the extension to $\lambda =1$ yields different
results in the two cases.
The self-consistent solution of (\ref{beta}) for $\lambda =1$ is
 plotted in Fig.~\ref{fig3}.
We see that
there is a substantial downturn renormalization of
$\beta$ in the region $J^{cr}_2 - J_2 \ll \lambda^2$,
where the self-consistent solution is in fact invalid. If instead, we
approximate
the critical value of $J_2$ from the region of intermediate $\beta$
(see Fig.~\ref{fig3}), we obtain  the larger $J^{cr}_2 \sim 2.3$,
which is in better agreement with numerical results.
On the other hand, the perturbative solution (with $\beta_0$ in the
subleading terms) gives
a correct description of the sublattice magnetization at
small $J_2$, shows no unphysical downturn renormalization near the
transition, and yields $J^{cr}_2 \sim 2.73 J_1$,
which is reasonably close to the numerical result. For all these reasons,
 we plotted the perturbative solution for $N_0$ in Fig.~\ref{figN_0}.

\subsection{Transverse susceptibility}

We will use a direct  way to obtain the
transverse susceptibility in the ordered phase, that is
we will  apply a homogeneous transverse magnetic field
 and compute the induced magnetization. For definiteness, we will assume
that $N_0$ is directed along the
$z-$axis,  and apply a magnetic field in
the $x-$direction.

For the calculations of the transverse magnetization,
 we found it convenient to introduce new Bose-operators as
linear combinations of the original
$a-$ and $b-$bosons:
\begin{equation}
s_i = \frac{a_i + b_i}{\sqrt{2}}~~~~~~~~~~~~~~~p_i = \frac{a_i -
b_i}{\sqrt{2}}
\end{equation}
In terms of these
new operators, the transformation to bosons, extended to large
$\lambda$, is
\begin{eqnarray}
M_{i}^{z} &=& s_{i}^{+}p_{i}+ p_{i}^{+}s_{i}, ~~~~~~~~~~~~~~
L_{i}^{z} = - \lambda^{-1/2} (c_{i}^{+}U_{i} +U_{i}c_{i}), \nonumber
\\
M_{i}^{x} &=& p_{i}^{+}c_{i} +c_{i}^{+}p_{i}, ~~~~~~~~~~~~~~~
L_{i}^{x} =   \lambda^{-1/2} (s_{i}^{+}U_{i} +U_i s_{i}), \nonumber
\\
M_{i}^{y} &=& -i (s_{i}^{+}c_{i}-c_{i}^{+}s_{i}), ~~~~~~~~~
L_{i}^{y} =   -i \lambda^{-1/2} (p_{i}^{+}U_{i} - U_i p_{i}),
\label{MLnew}
\end{eqnarray}
where $U_i = (1- \lambda(s_{i}^{+}s_{i}
+p_{i}^{+}p_{i}+c_{i}^{+}c_{i}))^{1/2}$.

The advantage of using this new form of the transformation is that
a magnetic field applied along $x$, only introduces a
condensate of the $p-$field.
As the expectation value of $M_x$ is obviously site-independent, the
$c-$ and $p-$field condensates should have the same momentum, i.e.,
 the condensate of $p$ should also have a momentum $\pi$.

Let us first discuss the mean-field results. In the
mean-field approximation, the
transverse magnetization per spin is $M_{\perp} = M_x /2 = <p>~<c> =
\lambda^{-1}~(\gamma_0
\beta_0)^{1/2}$, where we have introduced
$\gamma = \lambda {<p>}^2$, and $\gamma =\gamma_0$ at the mean-field
level.
 The mean-field ground state energy depends on both, $\beta_0$ and
$\gamma_0$, and is given by
\begin{equation}
\lambda E_0 = J_2 (\beta_0 + \gamma_0) - 4J^*_1 \beta_0 (1 - \beta_0)
+
 8 J^*_{1}~\beta_0 \gamma_0 - 2 H_x ~(\gamma_0~\beta_0)^{1/2},
\label{EH}
\end{equation}
where $E_0$ is the energy per a pair of spins, and, as before, the
magnetic field is measured in units of $g\mu_{B}/\hbar$.
Differentiating over  $\gamma_0$ and substituting the result into
$M_{\perp}$,
 we obtain
\begin{equation}
M_{\perp} = \frac{(\gamma_0~\beta_0)^{1/2}}{\lambda}
 = \frac{1}{\lambda}~\frac{\beta_0 H_x}{J_2 + 8 J^*_1 \beta_0}
\label{mm-f}
\end{equation}
To obtain the
susceptibility, we need $M_{\perp}$ only at vanishing magnetic field.
Substituting $\beta_0 = (4J^*_1 - J_2)/8J^*_1$ into
(\ref{mm-f}), we find
\begin{equation}
\chi_{\perp} = \frac{1}{4 J_1}~\frac{4J^*_1 - J_2}{8J^*_1}
\label{chimf}
\end{equation}
Observe that at $J_2 =0$, we recover the
classical spin-wave result $\chi_{\perp} = 1/8J_1$.
For the spin-stiffness we  obtain using (\ref{cswmf})
\begin{equation}
\rho_s = \frac{J_1}{4\lambda^2}~\frac{16(J^*_1)^2 -
J^2_2}{16(J^*_1)^2}
\label{rhomf}
\end{equation}
For $J_2 =0$, we again recover the classical spin-wave result.
In the opposite limit, $J_2 \approx 4J_1$, $\rho_s =
(J_1/\lambda^2)~
[(4 J^*_1 -J_2)/8J_1]$.
Finally, for the ratio $N^2_0/\rho_s$  we have $N^2_0/\rho_s = 1/J_1$
independent of $J_2$.

We now obtain the expression for $\chi_{\perp}$ to order $O(\lambda)$.
{}From (\ref{MLnew}) we have
$M_{\perp} =\lambda^{-1}\sqrt{\beta \gamma} + \Delta M_{\perp}$,
 where $\Delta M_{\perp} = N^{-1} \sum_{k \not = \pi} <c^{\dagger}_k~p_k>$.
To compute $\gamma$ and $\Delta M$ to first order in $\lambda$,
 we will need the
excitation spectra of quasiparticles in the presence of the field.
To obtain them,  we substitute the
transformation to bosons into the spin Hamiltonian and restrict our
calculations to the terms which are
quadratic in bosons. The fluctuations of the $s-$field are decoupled
from the
other two modes, while the fluctuations of the $c-$ and $p-$fields
are coupled
in the presence of a field. The computation of the quadratic form in
the $c-$
and $p-$bosons again requires some care as one needs to carefully examine
all terms
in the expansion of the square root in (\ref{MLnew}), keeping in mind
that
both $\gamma$ and $\beta$ are not small in $\lambda$.
Assembling the
contributions from all terms in the series, we obtain
\begin{eqnarray}
{\cal H} &=& E_0 + \sum_k A_s (k) ~s^{\dagger}_k s_k + \frac{B_s
(k)}{2}~(s^{\dagger}_k s^{\dagger}_{-k} + s_{k} s_{-k}) +
A_p (k) p^{\dagger}_k p_k + \frac{B_p
(k)}{2}~(p^{\dagger}_k p^{\dagger}_{-k} + p_{k} p_{-k}) + \nonumber
\\
&& A_c (k)~c^{\dagger}_k c_k + \frac{B_c
(k)}{2}~(c^{\dagger}_k c^{\dagger}_{-k} + c_{k} c_{-k}) + C (k)
(c^{\dagger}_k
p_k + p^{\dagger}_k c_k) + D (k) (p^{\dagger}_k c^{\dagger}_{-k} +
p_{k}
c_{-k})
\end{eqnarray}
where
\begin{equation}
\lambda E_0 = J_2 (\beta + \gamma) - 4J^*_1 \beta (1 - \beta) [ 1 - 2
\lambda
Z_1 (\beta)] - 2 H_x \sqrt{\beta \gamma} + 8 J^*_{1}
\beta \gamma (1 + \lambda/(8(1 - \beta)^2)
\label{Echi}
\end{equation}
and
\begin{eqnarray}
A_s (k) &=& J_2 + 2 J^*_1 \left (2 \beta  + \nu_k (1 - 2 \beta -
2\gamma)\right),~~~B_s (k) = 2 J^*_1 \nu_k (1 - 2\gamma) \nonumber \\
A_p (k) &=& J_2 + 2 J^*_1 \left (2 \beta +\frac{\beta
\gamma}{1-\beta} + \nu_k \left(1 - 2\beta -
\gamma + \frac{\beta \gamma}{1 - \beta}\right)\right), \nonumber \\
B_p (k) &=& 2 J^*_1 \left (\frac{\beta \gamma}{1 - \beta} - \nu_k
\left(1 -
\gamma - \frac{\beta \gamma}{1 - \beta} \right)\right) \nonumber \\
A_c (k) &=& J_2 + 2 J^*_1 \left (\beta \frac{4 - 3 \beta}{1 - \beta}
+
\frac{\beta^2 \gamma}{(1 - \beta)^2} + \nu_k \left(\frac{(1 -
2\beta)^2}{1-\beta} - \gamma \frac{2 - 4 \beta + \beta^2}{(1 -
\beta)^2}
\right)\right), \nonumber \\
B_c (k) &=& 2 J^*_1 \left (\beta \frac{2 - \beta}{1 - \beta} +
\frac{\beta^2 \gamma}{(1 - \beta)^2} + \nu_k \left(\frac{(1 -
2\beta)^2}{1-\beta} - \gamma \frac{2 - 4 \beta + \beta^2}{(1 -
\beta)^2}
\right)\right), \nonumber \\
C (k) &=& -H_x + 2 J^*_1~(\beta~ \gamma)^{1/2}~ \frac{3 - 2 \beta -
\nu_k (2 - 3 \beta)}{1 - \beta} \nonumber \\
D (k) &=& 2 J^*_1~(\beta~ \gamma)^{1/2}~\frac{1 -
\nu_k (2 - 3 \beta)}{1 - \beta}
\end{eqnarray}
Diagonalizing then the
$2\times 2$ matrix for $s-$bosons and $4\times 4$ matrix for coupled
$c-$ and
$p-$bosons by usual means, we obtain three branches of quasiparticle
excitations. The dispersion of the
$s-$boson has a gap,
 $\epsilon_s (\pi,\pi) \equiv H_x$. This result is
valid for any $J_2$ and is indeed the expected result
since $s$-quanta describe the fluctuations of the transverse
components of ${\vec M}$, and the $k=\pi$ mode of these fluctuations is
a homogeneous precession of the magnetization around the direction of the field
(we recall that both condensates have momentum $\pi$,
 and hence homogeneous ($k=0$) modes of composite fields
$M_z$ and $M_x$ correspond  to $k=\pi$ mode of the $s-$boson).
Furthermore, we found after a diagonalization that
one of the two coupled modes of
the $c-$ and $p-$bosons  remain gapless at $k=\pi$, while the other
has a gap which in the absence of the field is the same as the gap
for the $c-$quanta. The presence of a gapless excitation
 is a direct consequence of the Goldstone theorem.

For the calculation of the transverse magnetization,
we actually need only the ground state energy. Collecting the zero-point
contributions, which appear after diagonalization, we obtain
to order $H^2_x$
\begin{equation}
E_{tot} = E_0 - \frac{1}{2}~ \sum_i~\sum_k [A_i (k) - \epsilon_i (k)]
-
\sum_k l^2_p~l^2_c~\frac{[C (x_p + x_c) + D (1 + x_p
x_c)]^2}{\epsilon_p +
\epsilon_c}
\label{Etot}
\end{equation}
where $E_0$ is given by (\ref{Echi}),  $i = s,p$ or $c$, $\epsilon_i
= (A^2_i -
B^2_i)^{1/2}$, and $l^2_i = (A_i + \epsilon_i)/2 \epsilon_i,~x_i =
-B_i/(A_i +
\epsilon_i)$. Simultaneously, substituting old Bose operators in terms of
new ones into $\Delta M_{\perp}$, we also obtain
\begin{equation}
\Delta M_{\perp} = - \sum_k l^2_p~l^2_c~(x_p + x_c)~
\frac{C (x_p + x_c) + D (1 + x_p x_c)}{\epsilon_p +
\epsilon_c}
\label{dm}
\end{equation}
Evaluating $\gamma \propto \beta H^2_x$ from $\partial
E_{tot}/\partial
\gamma =0$ and using (\ref{beta}) and (\ref{dm}), we obtain the
result for
$\chi_{\perp} = M_{\perp}/H_x$ to order $O(\lambda)$. The full
expression is, however, too cumbersome to be presented here, so we analyze
only the limiting
cases and plot the result for arbitrary $J_2$ in
Fig.~\ref{figchi} (we used the same procedure of extending the result to
 $\lambda =1$ as for the sublattice magnetization).
Near the critical point, we found
\begin{eqnarray}
(\gamma ~\beta)^{1/2} &=& \frac{\beta H_x}{J^{cr}_2 - 5 \lambda J^*_1
Z_{\gamma}}
\nonumber \\
\Delta M_{\perp} &=& \frac{H_x \beta}{12 \pi \lambda J^*_1}~
\left(\frac{\lambda^2}{\beta}\right)^{1/2}~ \left(1 + O(\beta)\right)
\end{eqnarray}
where $Z_{\gamma} = N^{-1} \sum_k \nu_k/\sqrt{1 + \nu_k} = -0.328$,
$J^{cr}_2$
is given  by (\ref{eq21}), and we keep $\beta$ rather than $\beta_0$
in the
$O(\lambda)$ terms.
Collecting the two
 contributions to $M_{\perp}$, we obtain
\begin{equation}
\chi_{\perp}  = \frac{Z_{\chi}}{4J_1}~
\beta
{}~\left (1 + \frac{1}{3 \pi}~
\left(\frac{\lambda^2}{\beta}\right)^{1/2} +
O\left(\frac{\lambda^2}{\beta}\right) + \ldots\right)
\label{chirho}
\end{equation}
where $Z_{\chi} = 1 + 0.255\lambda$. The subleading term is a
Gaussian correction. Its divergence again implies that the self-consistent
approach only works for $\beta > \lambda^2$~\cite{co}.
In the opposite limit, $\lambda^2 \gg \beta$, a self-consistent theory
is inapplicable. Scaling considerations~\cite{CSY}
 predict that in this limit  $\chi_{\perp} = A
(Z_{\chi}/4J_1 a^2_0)~(\beta/\lambda^2)^{1/(1 +
\eta)}$, where $\eta \approx 0.03$ is the critical exponent for spin
correlations at criticality, and $A$ is a constant whose value cannot be
obtained in the present approach. At $\lambda^2 \gg \beta$, we also have
$\rho_s = \chi_{\perp} c^2_{sw} = A J_1
Z_{\rho}~(\beta/\lambda^2)^{1/(1 + \eta)}$, where
$Z_{\rho} = 1 - 0.257 \lambda$.

In the opposite limit, $J_2 =0$, we find
\begin{equation}
\chi_{\perp} = \frac{Z_{\perp}}{8J}
\end{equation}
where
\begin{equation}
Z_{\perp} =  1 - \lambda \frac{1}{N}~\sum_k \frac{\nu^2_k}{(1 -
\nu^2_k)^{1/2}}
{}~= 1 - 0.551 \lambda
\end{equation}
is the contribution from $s-$ and $p-$bosons, which is {\it exactly}
the same as in the first-order spin-wave theory. In other words,
longitudinal fluctuations do not contribute to the susceptibility
of a single-layer antiferromagnet. This is consistent with the fact
that the first-order $1/S$ result for $\chi_{\perp}$ agrees well with
the numerical data~\cite{singh}.
However, the longitudinal fluctuations are again small only for
$J_2/J_1 \ll 1$.
 As $J_2$ increases, our expression for $\chi_u$ deviates from
the spin-wave result, and eventually turns to zero much earlier than in
the self-consistent spin-wave theory.

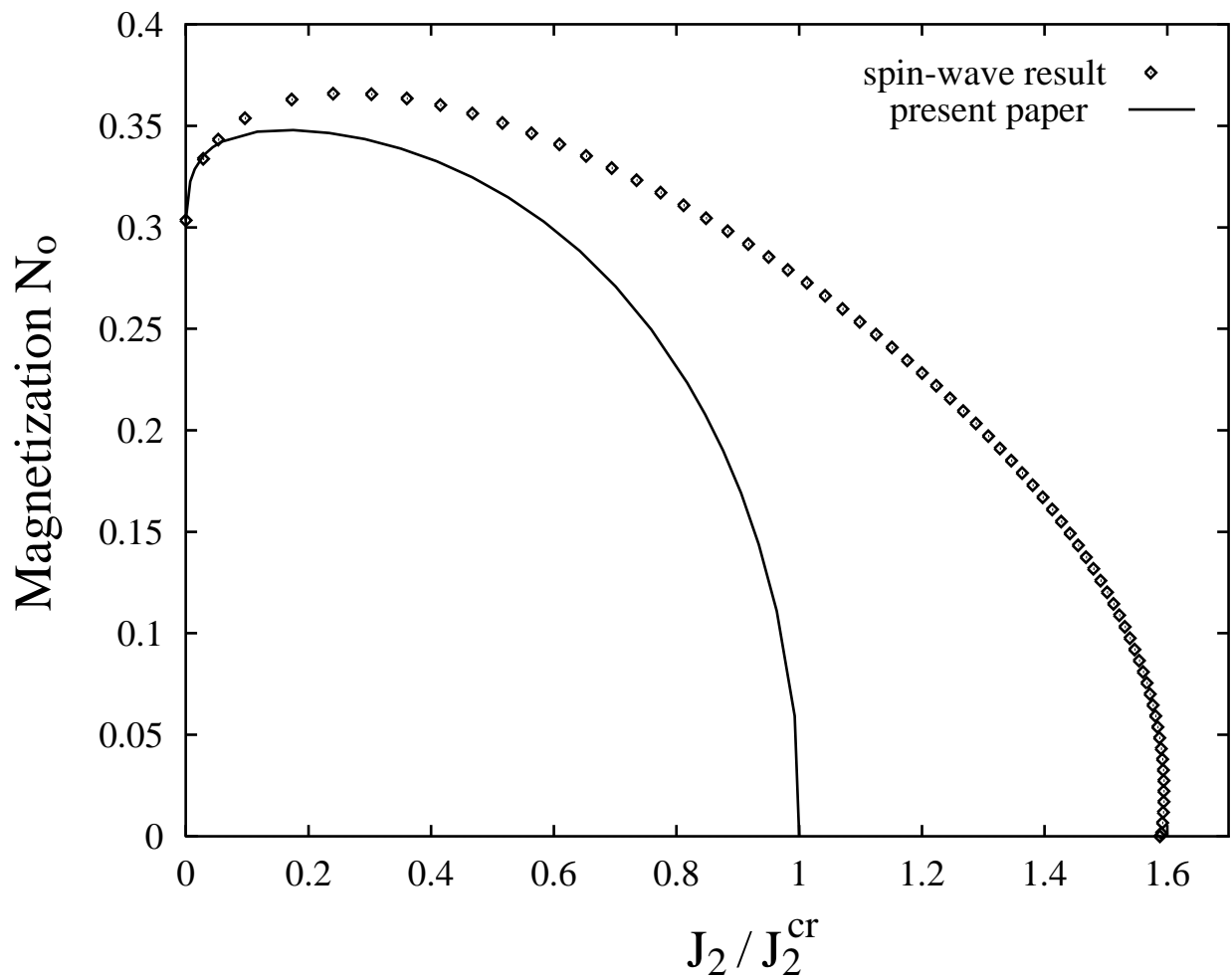
\begin{figure}
\caption{Sublattice magnetization as a function of $J_2/J_1$. Points -
the self-consistent spin-wave result; solid line -
the result of our present calculations which take longitudinal
spin fluctuations into account. The critical value of interlayer exchange is
$J^{cr}_2 = 2.73J_1$ (see appendix A).}
\label{figN_0}
\end{figure}

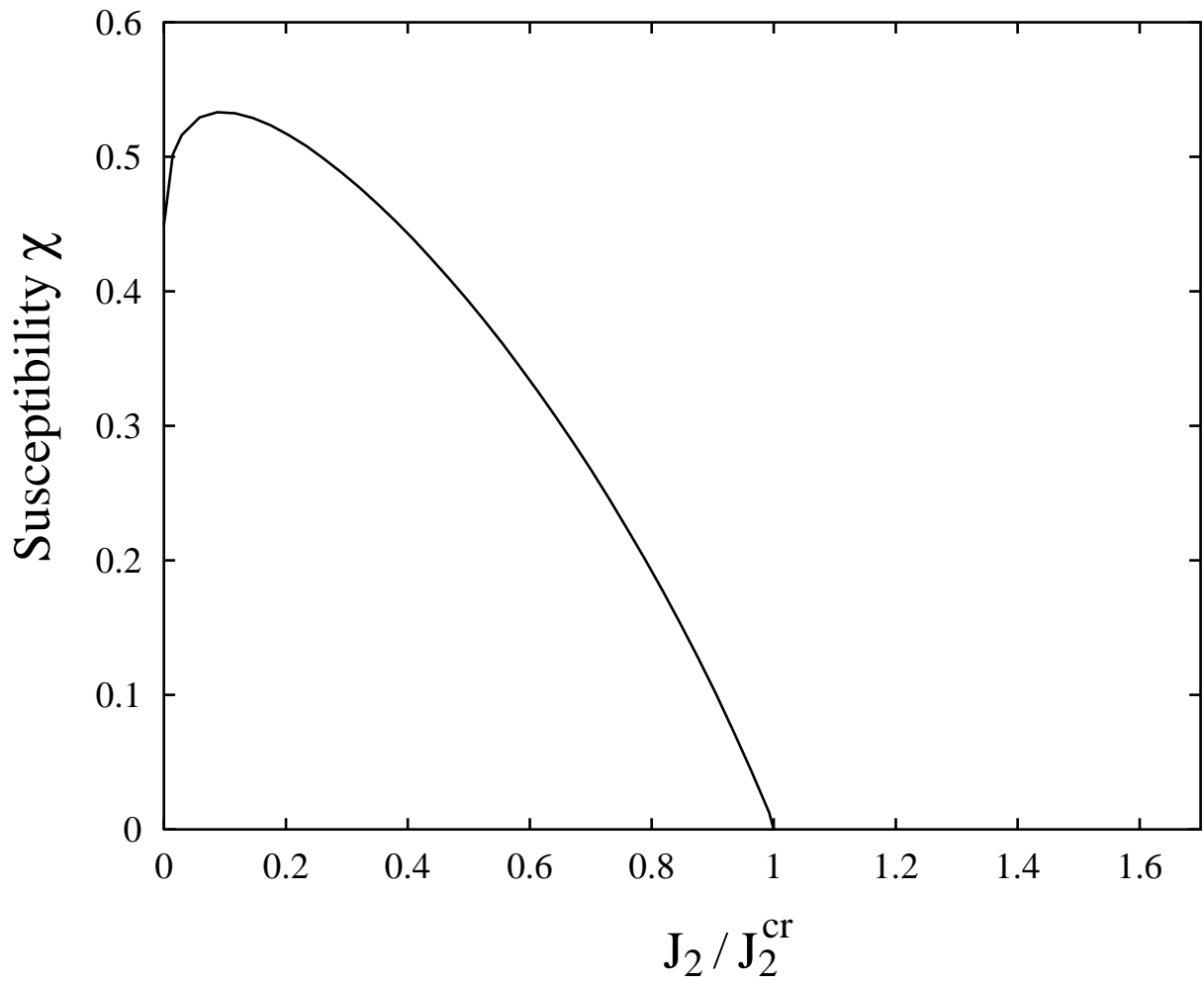
\begin{figure}
\caption{Transverse susceptibility in the ordered phase as a function of
$J_2/J_1$. The critical value of $J_2$ is the same as in
Fig.\protect\ref{figN_0}.}
\label{figchi}
\end{figure}

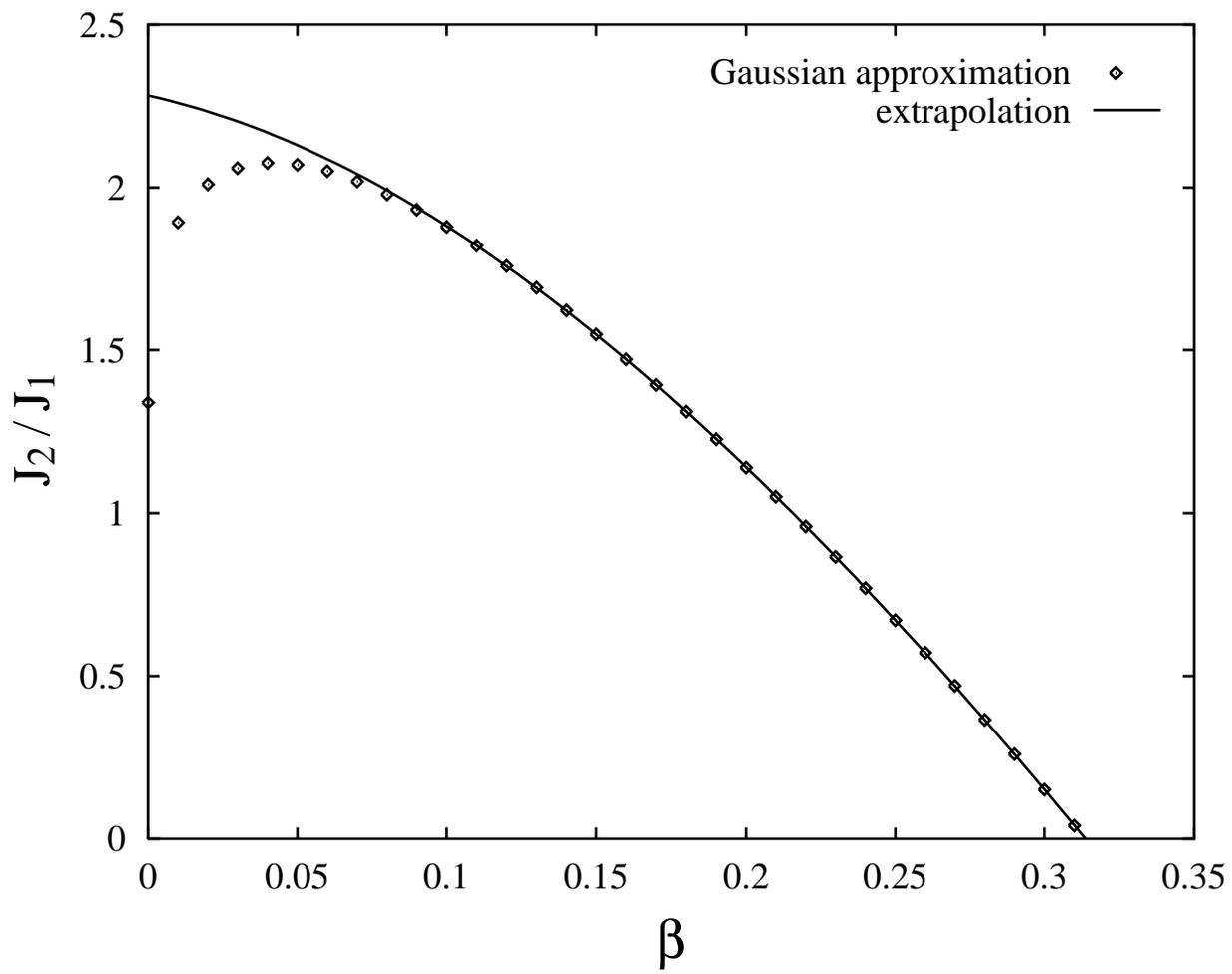
\begin{figure}
\caption{The solution of the self-consistent equation for
the fully renormalized value of the single-particle condensate, $\beta =
<c_{\pi}>^2$. Points are the
results of the self-consistent calculations
 extended to the physical case of $\lambda =1$.
The downturn renormalization
at small $\beta$ is due to divergent Gaussian fluctuations,
 and is probably unphysical. The solid line is the extrapolation of the
self-consistent formula at intermediate $\beta$ to $\beta =0$.}
\label{fig3}
\end{figure}

\begin{references}
\bibitem{co} Notice that the square-root divergence of the Gaussian
term
is consistent with the Josephson
relation $\chi_{\perp} \sim (\xi_J)^{D-2}$
where $D$ is a spacetime dimension. In the Gaussian approximation,
$\xi_J \propto \beta^{1/2}$, i.e., for $D=3$,
$\chi_{\perp} \propto \beta^{1/2}$ as in (\protect\ref{chirho}).
\end{references}
\end{document}